\documentclass[twocolumn]{aastex631}

\accepted{April 19, 2024}

\usepackage{amsmath}

\begin{document}

\title{
A Relativistic Formula for the Multiple Scattering of Photons
}

\correspondingauthor{Rohta Takahashi}
\email{takahashi@tomakomai-ct.ac.jp}

\author[0000-0002-3110-5732]{Rohta Takahashi}
\affiliation{National Institute of Technology, Tomakomai College, Tomakomai 059-1275, Japan}

\author{Masayuki Umemura}
\affiliation{Center for Computational Sciences, University of Tsukuba, Tsukuba 305-8577, Japan}

\author{Ken Ohsuga}
\affiliation{Center for Computational Sciences, University of Tsukuba, Tsukuba 305-8577, Japan}

\author{Yuta Asahina}
\affiliation{Center for Computational Sciences, University of Tsukuba, Tsukuba 305-8577, Japan}

\author{Rintaro Takeda}
\affiliation{Center for Computational Sciences, University of Tsukuba, Tsukuba 305-8577, Japan}

\author{Mikiya M. Takahashi}
\affiliation{Center for Computational Sciences, University of Tsukuba, Tsukuba 305-8577, Japan}
\affiliation{National Institute of Technology, Tokyo College, Hachioji 193-0997, Japan}

\author{Norita Kawanaka}
\affiliation{National Astronomical Observatory of
Japan (NAOJ), 2-21-1, Osawa, Mitaka, Tokyo 181-8588,
Japan}
\affiliation{Department of Physics, Graduate School of Science Tokyo Metropolitan University 1-1,
Minami-Osawa, Hachioji-shi, Tokyo 192-0397}
\affiliation{Center for Gravitational Physics and Quantum Information, Yukawa Institute for Theoretical Physics, Kyoto University, Kitashirakawa Oiwake-cho, Sakyo-ku,
Kyoto 606-8502, Japan}

\author{Kohkichi Konno}
\affiliation{National Institute of Technology, Tomakomai College, Tomakomai 059-1275, Japan}

\author{Tomoaki Nagasawa}
\affiliation{National Institute of Technology, Tomakomai College, Tomakomai 059-1275, Japan}

\begin{abstract}
We have discovered analytical expressions for the probability density function (PDF) of photons that are multiply scattered in relativistic flows, under the assumption of isotropic and inelastic scattering. 
These expressions characterize the collective dynamics of these photons, ranging from free-streaming to diffusion regions. 
The PDF, defined within the light cone to ensure the preservation of causality, is expressed in a three-dimensional space at a constant time surface. 
This expression is achieved by summing the PDFs of photons that have been scattered 
$n$ times within four-dimensional spacetime. 
We have confirmed that this formulation accurately reproduces the results of relativistic Monte Carlo simulations.
We found that the PDF in three-dimensional space at a constant time surface can be represented in a separable variable form. 
We demonstrate the behavior of the PDF in the laboratory frame across a wide range of Lorentz factors for the relativistic flow.
When the Lorentz factor of the fluid is low, the behavior of scattered photons evolves sequentially from free propagation to diffusion, and then to dynamic diffusion, where the mean effective velocity of the photons equates to that of the fluid. 
On the other hand, when the Lorentz factor is large, the behavior evolves from anisotropic ballistic motion, characterized by a mean effective velocity approaching the speed of light, to dynamic diffusion. 
%
\end{abstract}




\section{Introduction} \label{sec:intro}
The relativistic effects of radiation field are essential in a variety of astrophysical flow and jets, e.g., active galactic nuclei (AGNs), gamma-ray bursts (GRBs), supernovae (SNs). 
%
%
%
In a black hole (BH) accretion flow with super-Eddington accretion rate, the photons and the fluid are tightly coupled and the radiative transfer (RT) of the scattered photons in a photon-trapping region is of great significance \citep{ohsuga2007supercritical,ohsuga2009global,mckinney2014three,takeo2020hyper,liska2022formation}. 
In order to investigate the dynamical radiation effects in a relativistic flow, some past studies tried to directly solve the RT equation, i.e. Boltzmann equation, for  photon \citep{beloborodov2011radiative,jiang2016iron,ohsuga2016numerical,takahashi2017general,asahina2020numerical,asahina2022general,takahashi20223d} and neutrino \citep[e.g.,][]{nagakura2018simulations,akaho2021multidimensional,akaho2023protoneutron}.
%

%
Despite the significant efforts of the past studies, even for the simplest cases assuming isotropic and elastic scattering of photons, the time dependent radiation field of photons (or neutrinos) in a relativistic flow has not been hitherto solved exactly. 
In a relativistic flow, the relativistic boosting effect is significant in the diffusion process \citep{krumholz2007equations,shibata2014random}. 
In particular, a precise description of the intermediate state between the free-propagating state and the diffusion state has not been possible to date, and lack of analytical understanding has prevented a fundamental understanding of these phenomena. 
In fact, we have confirmed that some of the approximate formulas proposed in the past contain a violation of the law of causality. 
Additionally, the relativistic diffusion problem is a well-known unsolved problem that has not been solved for many years \cite[e.g.,][]{dunkel2007relativistic}. The PDF provided in this paper, describing relativistic diffusion for photons, likely represents the first instance of an analytical solution for the relativistic diffusion problem involving these particles. 

In this paper, we shall pursue the analytic approach to describe the collective behavior of the repeatedly scattered photons in a relativistically boosted medium, and we investigate the time evolution of the scattering photons in the relativistic flow both in the rest frame (\S 2) and in the laboratory frame (\S 3). In the final section (\S 4), the conclutions are given. 

%
%


\section{Distribution of scattering photons in a static medium}

According to relativistic kinetic theory  \citep[e.g., ][]{lindquist1966relativistic,ehlers1971general,israel1972relativistic,sachs1971kinetic}, the collective behavior of particles is described by an invariant distribution function $\mathcal{F}(x^\mu,p^\mu)$ where $x^\mu$ and $p^\mu$ are, respectively, the coordinate and the momentum of a particle. 
%
%
%
The particle number density flux $N^\mu$ is given by $N^\mu=\int dN^\mu = \int \mathcal{F}(x^\mu, p^\mu)p^\mu dP$ and the particle number $N(x^\mu)dV$ in a volume element $dV$ at a time slice surface (where $t=t_0$) is given by the projection of $N^\mu$ onto the vector volume element, 
i.e., $N(x^\mu)dV
=dV\int \mathcal{F}(x^\mu, p^\mu)(-\hat{u}_\mu p^\mu)dP$ where $\hat{u}_\mu$ is a timelike unit vector. 
%
In this study, the probability density function (PDF) in a time slice surface are used to describe the collective behavior of scattered photons.
The PDF $P(x^\mu)|_{t=t_0}$ giving the probability that particles exist in some region at the time slice surface by volume integration is given by  $P(x^\mu)|_{t=t_0}=N(x^\mu)/N_{\rm all}$ where $N_{\rm all}$ is the number of photons in three space $V$ at $t=t_0$ calculated as $N_{\rm all}=\int_{V_{t=t_0}}N(x^\mu) dV$. 
Then, the PDF $P(t,r)$ satisfies the normalization $\int_{V_{t=t_0}} P(x^\mu)dV=1$.
The time component of $N^\mu(x^\mu)$ as $N^0(x^\mu)=N_{\rm all}P(x^\mu)$. 

We describe the coordinates $x^\mu$ and momentum $k^\mu$ of a photon as $x^\mu = (ct, \boldsymbol{r})$ and $k^\mu = (\hbar \omega/c, \hbar \boldsymbol{k})$ where $c$ is the speed of light and $\hbar$ is the Planck constant divided by $2\pi$. 
%
%
%
It is useful to introduce the normalization with the mean free path of photon $\ell_0$ in the fluid rest frame \citep{rybicki1991radiative,mihalas2013foundations}. 
Here, we introduce 
$x^\mu_* =(t_*, \boldsymbol{r}_*) \equiv x^\mu/\ell_0 =(ct/\ell_0, \boldsymbol{r}/\ell_0)$ and $k^\mu_* =(\omega_*, \boldsymbol{k}_*) \equiv k^\mu \ell_0 =(\hbar\omega\ell_0/c, \hbar\boldsymbol{k}\ell_0)$.  
%
%
%
We also define $r_* \equiv |\boldsymbol{r}_*|$ and $k_* \equiv |\boldsymbol{k}_*|$. 
The PDF, $P(t_*,\boldsymbol{r}_*)$, of a scattered photon is a function where $P(t_*,\boldsymbol{r}_*)d\boldsymbol{r}_*$ gives the probability that the scattered photon is in spatial region $d\boldsymbol{r}_*$ in the spatial hypersurface at time $t_*$.
These scattered photons include photons that have experienced $n$ times scatterings where $n=0, 1, 2, \cdots$. 
We have found the PDF, $P(t_*,\boldsymbol{r}_*)$, is expressed as the sum 
$P(t_*, \boldsymbol{r}_*)=\sum_{n=0}^\infty P_n(t_*, \boldsymbol{r}_*)$  
where $P_n(t_*, \boldsymbol{r}_*)$ is a PDF in four-dimensional spacetime that represents the distribution of the next scattering point of photons scattered $n$ times, which satisfy the normalization $\int P_n(t_*, \boldsymbol{r}_*) d^4x_* = 1$ where $d^4x_*$ is an invariant volume element of four-dimensional spacetime.  
\footnote{
It is noted that the sum $P(t_*, \boldsymbol{r}_*)=\sum_{n=0}^\infty P_n(t_*, \boldsymbol{r}_*)$ corresponds to the series expansion of the function $P(t_*, \boldsymbol{r}_*)$ in terms of scattering number $n$. 
Specifically, the analytic solution derived below satisfies the normalization $\int P_{n}(t_*, \boldsymbol{r}_*)d^3\boldsymbol{r}_*=e^{-t_*}t_*^n/n!$ $(n=0, 1, 2, \cdots)$ and the sum of the normalization is calculated as $\sum_{n=0}^\infty e^{-t_*}t_*^n/n!=1$ which corresponds to the sum of a Poisson distribution. 
}
In the following, we report the method and results of analytic calculation of $P_n(t_*, \boldsymbol{r}_*)$ under the assumptions of isotropic elastic scattering. 

Consider a process in which, at time $t_*=0$, a large number of photons are instantaneously emitted isotropically from a point in a static medium and spread out while repeatedly elastic scattering in the medium. 
Define the point of the photon emission as the coordinate origin, denote the position vector in spatial coordinate as $\boldsymbol{r}_*$ and the distance from the origin as $r_* (=|\boldsymbol{r}_*|)$. 
In this case, the PDF $P(x^\mu_*)$ at a time $t_*$ for the scattered photons exhibits a spatially spherical symmetric distribution. 
Consequently, the PDF is a function of $t_*$ and $r_*$, i.e., $P(t_*, r_*)$ and $P_n(t_*,r_*)$. 
The equation to be solved to derive the analytical formula for $P_n(t_*,r_*)$ is obtained by a method similar to that used in previous studies of random flight \citep[e.g.,][]{rayleigh1919xxxi,hughes1995random}. 

\begin{figure}[t]
\includegraphics[width=1.\columnwidth]{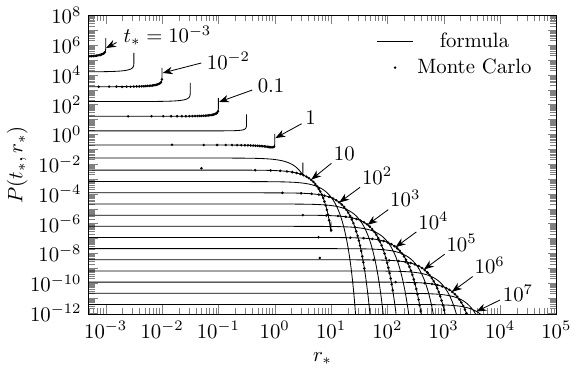}
\caption{
The probability density function $P(t_*,r_*)$ of scattering photons in a static medium at the time $t_*=10^{-3}$, $10^{-2.5}$, $10^{-2}$, $10^{-1.5}$, $10^{-1}$, $10^{-0.5}$, $1$, $10^{0.5}$, $10$, $10^{1.5}$, $10^{2}$, $10^{2.5}$, $10^{3}$, $10^{3.5}$, $10^{4}$, $10^{4.5}$, $10^{5}$, $10^{5.5}$, $10^{6}$, $10^{6.5}$ and $10^{7}$ ({\it from top to bottom}). The results of the formula ({\it solid line}) and the Monte-Carlo simulations ({\it blank circle}) are shown. 
\label{fig:Prest}}
\end{figure}

\begin{figure*}[ht!]
\begin{center}
\includegraphics[width=1.7\columnwidth]{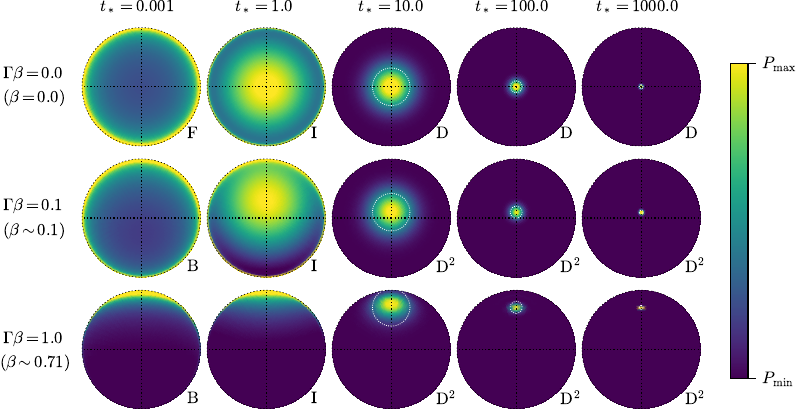}
\end{center}
\caption{
The probability density function $P(t_*,r_*)$ of scattering photons in the laboratory frame of a relativistic flow with $\Gamma\beta=0.0$ ({\it the first column}), $0.1$ ({\it the second column}) and $1.0$ ({\it the third column}) at the time $t_*=0.001$, $1.0$, $10.0$, $100.0$ and $1000.0$ ({\it from left to right}). In each plot, the  abscissa and the ordinate represent $x_*$-axis and $z_*$-axis, respectively, and the outermost circle ({\it dotted circle}) shows the location of $r_*=t_*$. The symbols (F, B, I, D, D$^2$) at the bottom-right corner of each plot represents the state of the scattering photons (see, texts). The maximum and minimum values, $P_{\rm max}$ and $P_{\rm min}$, of the color bar are set to show the distribution of the probability density function. The circle with the radius of $\sqrt{t_*}$ and the center at $(x_*,z_*)=(0,\beta t_*)$ are shown at the time $t_*=10.0$, $100.0$ and $1000.0$ ({\it white dotted circle}).
\label{fig:Plab}}
\end{figure*}

If we set the initial position of a photon at the origin of the coordinates,  the probability $p_0(x^\mu_*)$ is given by the delta function as $p_0(x^\mu_*)=\delta(x^\mu_*)$, and its Fourier transform is $\hat{p}_0(k_*)=1$. 
We denote the coordinate $x^\mu_*$ describing $P_n$ as $x_{*n}$.  
The relation between $P_n(x_{*n})$ and $P_{n-1}(x_{*n-1})$ is described by the convolution given as $P_{n}(x_{*n}) = \int p(x_{*n}-x_{*n-1}) P_{n-1}(x_{*n-1}) d^4 x_{*n-1}$, 
where $p(x_{*n}-x_{*n-1})$ is a probability that a photon is scattered from $x_{*n-1}$ to $x_{*n}$. 
%
%
The PDF of an isotropic scattering $p(t_*, \boldsymbol{r}_*)$ is given by $p(t_*, \boldsymbol{r}_*) = p(t_*, r_*) = \delta(t_*-r_*)\theta(t_*)e^{-r_*}/(4\pi r_*^2)$, and its 
Fourier transform is $\hat{p}(\omega_*, k_*) = \arctan [k_*/(1-i\omega_*)]/k_*$. 
Since the convolution theorem tells $\hat{P}_{n}(k_*) = \hat{p}(k_*) \hat{P}_{n-1}(k_*)$ \citep[e.g.,][]{folland2009fourier},   we obtain $\hat{P}_n(k_*) = \hat{p}(k_*)^{n+1} \hat{p}_0(k_*)=\hat{p}(k_*)^{n+1}$.  
Then, the PDF $P_n(t_*, r_*)$ is calculated by the inverse Fourier transform given by $P_n(t_*, r_*)=(2\pi)^{-4}\int e^{ik_{*\mu} x_*^\mu}\{\hat{p}(\omega_*, k_*)\}^{n+1} d^4k_*$. 
%
Finally, we can obtain the variable separation form of $P_n(t_*, r_*)$ given by $P_n(t_*, r_*) = t_*^{n-3} e^{-t_*} V_n(r_*/t_*)\theta(t_*)/(4\pi)$ 
where $\theta(t_*)$ is a step function 
and $V_n(r_*/t_*)$ is an even function of $r_*/t_*$.
Since the function $P_n(t_*, r_*)$ 
contains $r_*$ in the form $r_*/t_*$, the function $P_n(t_*, r_*)$ is in variable separation form of variables $t_*$ and $v\equiv r_*/t_*$.
It should be noted that the function $V_n(v)$ defined in the range $0\le v\le 1$, i.e., $0\le r_*\le t_*$, ensures that the formula for the PDF preserves causality.
The derivation of the analytical expression for the function $V_n(v)$ is explained below.

For $n=0$, $P_0(t_*,r_*)=p(t_*,r_*)$ as denoted above. 
We can find the function $V_n(v)$ obeys the equation $v V_n(v) = S_n (v)$ for $n=1$ and the differential equations 
$d^{n-1}\{vV_n(v)\}/dv^{n-1}=S_n(v)$ for $n\ge 2$  
where $S_n(v)=(i^{n-2}/\pi)\{f_n(-iv)-(-1)^{n+1} f_n(iv)\}$. 
Here, $f_n(x)\equiv (\pi/2-\tan^{-1}x)^{n+1}$. 
The solution of 
$V_n(v)~(n\ge 2)$ is described as 
$V_2(v)=W_2(v)/v$ and 
$V_n(v)=\sum_{k=0}^{k_{\rm max}}C_{n}^k v^{2k}+W_{n}(v)/v$ (for $n\ge 3$) 
where $C_n^k$ are constants, $W_n(v)$ is a function satisfying $W_n(0)=0$, and $k_{\rm max}=n/2-2$ ($(n-1)/2-1$) for even (odd) integer $n$. 
%
Using the properties of the function $P_n(t,r)$, the coefficient $C_n^k$ is computed as 
\begin{equation}
C_{n}^k = \frac{2(-1)^k}{\pi(2k+1)!(n-3-2k)!}
\int_0^\infty \frac{(\tan^{-1}X)^{n+1}}{X^{n-1-2k}}dX,
\label{eq:Cnk}
\end{equation}
which can be analytically solved by integration by parts. 
With a new variable $y=2\tanh^{-1}v$, the function 
$W_{n}(v)$ can be solved by $n-1$ times integration by parts and we obtain the expression $W_{n}(v) = \sum_{k=0}^{n+1} c_{n+1}^k I_{n-1}^k(y(v))$ where 
the coefficients $c_n^k$ is given as  
$c_n^k=(i\pi)^{n-k-1}2^{-n}{}_n C_k \{(-1)^n-(-1)^k\}$ 
%
%
and $I_n^k(y)$ is solved by the recurrence relation given as 
\begin{eqnarray}
I_n^k(y) &=& \bigg[
    I_{n-1}^k(\bar{y})s_0^1(\bar{y})h(\bar{y})
    +\sum_{i=1}^{n-2}(-1)^i s_1^i(\bar{y})I_{n-1-i}^k(\bar{y})
    \nonumber\\&&
    +\sum_{i=0}^k (-1)^{i+n+1}s_{i+1}^{n-1}(\bar{y})\frac{d^i}{d\bar{y}^i} \bar{y}^k
    \bigg]_{\bar{y}=0}^{\bar{y}=y},
\end{eqnarray}
where $h(y)=1+\cosh y$ and the function $s_n^m(y)$ is defined by 
$s_0^1(y)=\tanh(y/2)/h(y)$, 
$s_n^m(y)=\int_0^y s_{n-1}^m (y_1)dy_1$ and 
$s_0^m(y)=s_1^{m-1}(y)/h(y)$. 
%
The coefficients $C_n^k$ and the functions $s_n^m(y)$ can be solved analytically by Mathematica to obtain an analytical expression for the function $V_n(v)$.
%
We obtain the analytic results of $V_n(v)$ as, e.g., 
$V_1(v)=(2/v)t(v)$, 
$V_2(v) = -(\pi^2/4)f(v)+3f(v)t(v)^2-(3/v)f_2(v)+(6/v)f_+(v)t(v)$, 
and 
\begin{eqnarray}
V_3(v) &=& -\frac{\pi^2}{2}f(v)-\frac{\pi^2(v^2+1)}{2v}t(v) +6f(v)t(v)^2 \nonumber\\&&
    +2v f(v)^2 t(v)^3 +\frac{\pi^2}{2}\{f_+(v)+f_-(v)\} \nonumber\\&&
    +\frac{12}{v}f_+(v)t(v) -12f_+(v)t(v)^2 +6f_3(v)
    \nonumber\\&&
    +6(2t(v)-v^{-1})f_2(v), \nonumber
\end{eqnarray}
where $t(v)=\tanh^{-1}v$, $f(v)=1/v-1$, $f_\pm(v)=\log 2/(1\pm v)$, $f_n(v)={\rm Li}[(v-1)/(v+1)]$ where ${\rm Li}_n$ is the polylogarithm of order $n$. 
%
%
Similarly, we obtained analytical expressions for $V_n(v)$ ($n=4, 5, \cdots, 10$), but as $n$ grows, the expressions become longer and it is not practical to calculate $P(t_*,r_*)$ using these closed form of the analytic expression of $V_n(v)$. 
Instead, when we compute $P(t_*,r_*)$, we use the series expansion of $V_n(v)$ for large $n$ given as $V_n(v)=\sum_{k=0}^\infty C_n^k v^{2k}$ where the coefficients of the series expansion are calculated by Equation (\ref{eq:Cnk}) and the expansion coefficient of $S_n(v)$ which can be analytically obtained. 

When we calculate $P(t_*, r_*)$ from the sum of $P_n(t_*, r_*)$, we should set the range of $n$. 
If $t$ is large ($t_*\gtrsim 10^2$), since the major portion of $P_n$ resides in the range of $n-\sqrt{n}<t_*<n+\sqrt{n}$, we set the range of n as $t_*-c_*\sqrt{t_*}<n<t_*+c_*\sqrt{t_*}$ where the value of $c_*$ is set to achieve sufficient numerical precision ($c_*\gtrsim 3$). 
When $t_*$ is much larger ($t_*>10^3$), $P$ can be treated as a perturbation of the diffusion approximation $P_{\rm diff}$ (defined below). 
In this case, the expansion coefficients of $P$ are obtained by extrapolation with the pre-computed $P$ at some time $t_*$ and $P_{\rm diff}$. 
In this study, we use 20th order interpolation for this calculation.
On the other hand, when $t_*$ is not large ($t_*\lesssim 10^2$), the sum of $P_n$ can be calculated directly. 

%
%
Figure \ref{fig:Prest} shows the PDF of scattering photons in a static medium from the time $t_*=10^{-3}$ to $10^7$. 
For $t_*\lesssim 10$, the spike-like peak can be seen at $r_*=t_*$, which corresponds to the peaks held by $P_1(t_*,r_*)$ and $P_2(t_*,r_*)$.
In other words, this peak represents the traces of photons propagating at the speed of light. 
We call this period the isotropic free-propagation state (F).
On the other hand, in case $t_*\gtrsim 10$, the maximum of the distribution lies at the center of $r_*=0$. 
As time passes ($t_*\gg 1$), the PDF $P(t_*, r_*)$ of the scattering photons approaches the distribution $P_{\rm diff}(t_*, r_*)$ of the diffusion approximation, i.e., $P_{\rm diff}(t_*, r_*)=\{3/(4\pi t_*)\}^{3/2}\exp\{-3r_*^2/(4t_*)\}$.
Thus, we call this period the diffusion state (D).
The period around $t_*\sim 1 \sim 10$ is in the intermediate state (I) between the two states (F and D). 
That is, it has a local maximum of the PDF at the center of $r_*=0$ and also a spike-like peak at $r_*=t_*$. 
Results of Monte-Carlo (MC) simulations are shown as dots in Figure \ref{fig:Prest}, confirming that our analytic formulas reproduce the simulation results.
In the calculations in Figure \ref{fig:Prest}, using the analytical solution can accelerate the computation by a factor of $10^2\sim 10^5$ compared to MC simulations. In addition, in the case of MC, it is difficult to accurately obtain values near the maximum of the PDF. Therefore, the analytical solution is superior to the MC calculation in terms of computation time, computational accuracy, and computational feasibility. 


\section{Distribution of scattering photons in a relativistic flow}

\begin{figure}[t]
\includegraphics[width=1.0\columnwidth]{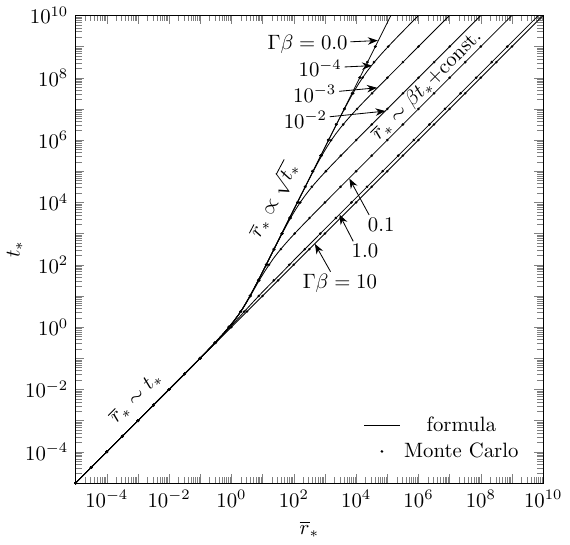}
\caption{
The time evolution of the mean radial distance $\overline{r}_*$ of the probability distribution of the scattering photons in the laboratory frame of the fluid with $\Gamma\beta=0.0$, $10^{-4}$, $10^{-3}$, $10^{-2}$, $0.1$, $1.0$ and $10$ ({\it from top to bottom}). The results of the formula ({\it solid line}) and the MC simulations ({\it blank circle}) are shown. The formula representing the state of the scattering photons are shown: $\overline{r}_*\sim t_*$ (isotropic free-streaming or anisotropic ballistic motion), $\overline{r}_*\propto \sqrt{t_*}$ (diffusion, i.e., diffusive expansion) and $\overline{r}_*\sim \beta t_*+$const. (dynamic diffusion, i.e., boosted diffusion). 
\label{fig:rmean}}
\end{figure}

We can also calculate the photon number density flux $N^\mu$ in the rest frame analytically.
The time component of this flux is $N^0(x^\mu_*)=N_{\rm all} P(x^\mu_*)$ and the flux satisfies the particle number conservation law, $\nabla_\mu N^\mu$=0.
Then, the spatial component of the flux $N^i$ ($i=1, 2, 3$) is calculated as 
$N^i = N_{\rm all} Q(t_*,r_*)$ 
where $Q(t_*,r_*)$ is defined as 
\begin{equation}
Q(t_*,r_*) = \frac{r_*}{t_*} \left\{ P(t_*,r_*)+\sum_{n=0}^\infty c_{n}(t_*)X_{n}\left(\frac{\sqrt{n}r_*}{t_*}\right) \right\}
\end{equation}
Here, $c_{n}(t_*)=n^{3/2}t_*^{n-3}e^{-t_*}(t_*-n)/n!$ 
and 
$X_{n}(u)\equiv u^{-3}\int_0^u \tilde{u}^2U_{n}(\tilde{u})d\tilde{u}$ 
where $U_{n}(u)\equiv V_n(u/\sqrt{n})n!/(4\pi\sqrt{n^3})$. 
The function $X_n(u)$ can be calculated analytically as the closed analytic form 
or as the series expansion form as $X_n(u)=\sum_{k=0}^\infty C_k u^{2k}/(2k+3)$ when $U_n(u)$ is expanded as $U_n(u)=\sum_{k=0}^\infty C_k u^{2k}$.  

The photon number density flux in the laboratory frame $N^\mu$ is calculated as $N^\mu=\Lambda^\mu_\nu {N'}^\nu$ where $\Lambda^\mu_\nu$ is a matrix representing the Lorentz transformation and ${N'}^\nu$ is the photon number density flux in the rest frame. 
The PDF in the laboratory frame is calculated as $P(x_*^\mu)=N^0/N$ where $N^0$ is the time component of the photon number density flux and $N$ is the number density at a constant time surface in the laboratory frame which is calculated by the integration of $N^0$ in three dimensional space in the laboratory frame. 

Figure \ref{fig:Plab} shows the cross-section in the $x_*$-$z_*$ plane of the PDF calculated by the analytic formula in the laboratory frame. 
Figure \ref{fig:rmean} shows the time evolution of the mean of $r_*$, $\overline{r}_*$, in the laboratory frame defined by the following three-dimensional integration in space, 
$\overline{r}_*(t_*)=\int_{\rm 3d\,space} r_*\, P(t_*,\boldsymbol{r}_*) d^3\boldsymbol{r}_*$, 
%
where $P(t_*,\boldsymbol{r}_*)$ is the PDF at the constant $t_*$ surface in the laboratory frame. 
Figure \ref{fig:rmean} shows the results of both the analytical formula ({\it solid line}) and the MC simulation ({\it black circle}).

In Figure \ref{fig:Plab}, the case of $\Gamma\beta=0$ ({\it the first column}) reflects the distribution described in Figure \ref{fig:Prest}. 
For $\Gamma\beta=0$, approximately, $\overline{r}_*=t_*$ (i.e., light-speed propagation) for $t_*\lesssim t_a$, and $\overline{r}_*= 4\sqrt{t_*/(3\pi)}$ (diffusive expansion) for $t_*\gtrsim t_a$, where the time $t_a$ is the boundary between the free propagating and diffuse states, calculated as $t_a=16/(3\pi)$.
In Figure \ref{fig:rmean}, we can confirm that the analytical formulas and the Monte Carlo simulation results are consistent for all periods.

As shown in Figure \ref{fig:Plab}, for the other cases, $\Gamma\beta=0.1$ ({\it the second column}) and $1.0$ ({\it the third column}), the distribution of the PDF is biased in the $+z_*$-direction due to the boost effect at all the times. 
For $t_*\lesssim 1$, we see that most of the photons are concentrated near $z_*=t_*$. 
That is, for $\Gamma\beta>0$ and $t_*\lesssim 1$, they propagate at the speed of light anisotropically in the boost direction, in contrast to the isotropic propagation at the speed of light for $\Gamma\beta=0$. 
When $t_*\lesssim 1$, we can see the anisotropic ballistic motion [denoted as (B) in Figure \ref{fig:Plab}] whose mean velocity is nearly equal to the speed of light. 
Figure \ref{fig:rmean} also confirms that the propagation is at the speed of light when $t_*\lesssim 1$. 
For $t_b\lesssim t_*$, the scattered photons are in a state of dynamic diffusion [denoted as (D$^2$) in Figure \ref{fig:Plab}], where the time $t_b$ is the boundary between the diffusion and the dynamic diffusion and is calculated as $t_b=(4/(3\pi\beta^2))[1+(1+2\beta^2/\Gamma)^{1/2}]^2$. 
That is, as can be seen in Figure \ref{fig:Plab}, they expand diffusively while being boosted in the $z_*$-direction. 
In this case, the center of the distribution moves approximately according to $\overline{r}_*=\beta t_*-8\beta/(3\pi\Gamma)$. 
This can be seen in both Figure \ref{fig:Plab} and Figure \ref{fig:rmean}. 
In Figure \ref{fig:Plab}, when $\Gamma\beta\gtrsim 0.1$, we can see the intermediate state (I) between the two states (B and D$^2$) around $t_*\sim 1$. 

\section{Conclusions}
%
In this study, under the assumption of isotropic and elastic scattering, we analytically describe the PDF $P(x_*^\mu)$ in a time slice surface and the PDF $P_n(x_*^\mu)$ in spacetime that describe the collective behavior of scattering photons in a relativistic fluid.
The derived equations reproduce the results of the MC simulations and succeeds in smoothly describing the intermediate state between the free-streaming (F) and the diffusion (D) state when $\Gamma\sim 0$ or between the anisotropic ballistic motion (B) and the dynamic diffusion (D$^2$). 
Although we considered a point source represented by the delta function in this study, we believe that the distribution of photons emitted from spatially spread out or temporally continuous radiation sources can be represented by the superposition of the present analytical solution. 
In the future, the authors intend to implement this study in the general relativistic radiative transfer simulations and extend it to more general case of photon diffusion. 
Based on the present study, some of the authors of this paper are working on calculations of the distribution function $\mathcal{F}(x_*^\mu, k_*^\mu)$ in phase space for the scattering photons. 


\begin{acknowledgments}
We are deeply grateful to the anonymous referee for their insightful and constructive comments, which have significantly improved the quality of the manuscript. The authors are grateful to H. Itoh, H. Takahashi and T. Kawashima for valuable discussions. This work was supported by JSPS KAKENHI Grant numbers 15H03638 (M.U.), 16K05302 (R.T., M.U.), 22J10256 (M.M.T.), 18K03710, 21H04488 (K.O.), 19H00697 (M.U.and R.T.), 21H01132 (R.T., M.U., K.O. and Y.A.), 18K13581, 23K03445 (Y.A.), 22KJ0382 (M.T.), 22K03686 (N.K.). This research was also supported in part by Multidisciplinary Cooperative Research Program in CCS, University of Tsukuba. This work was also supported by MEXT as “Program for Promoting Researches on the Supercomputer Fugaku” (Structure and Evolution of the Universe Unraveled by Fusion of Simulation and AI, JPMXP1020240219; KO, Black hole accretion disks and quasi-periodic oscillations revealed by general relativistic hydrodynamics simulations and general relativistic radiation transfer calculations, JPMXP1020240054; KO), and by Joint Institute for Computational Fundamental Science (JICFuS, K.O.).
\end{acknowledgments}

\bibliography{ref}{}
\bibliographystyle{aasjournal}



\end{document}